\documentstyle[aps]{revtex}

\begin{document}
\title{Addendum on rotational properties of confined bosons: time-inversion
referencing}
\author{F. Brosens\thanks{%
Senior Research Associate of the FWO-Vlaanderen{}}, and J. T. Devreese%
\thanks{%
Also at the Universiteit Antwerpen (RUCA){}, and Technische Universiteit
Eindhoven, Eindhoven, The Netherlands.}}
\address{{Departement Natuurkunde, }\\
{Universiteit Antwerpen (UIA), Universiteitsplein }1, \\
B-2610 Antwerpen}
\author{L. F. Lemmens\thanks{%
e-mail lcnlmmns@ruca.ua.ac.be}}
\address{Departement Natuurkunde, \\
Universiteit Antwerpen (RUCA), Groenenborgerlaan 171, \\
B-2020 Antwerpen}
\maketitle

\begin{abstract}
In this addendum we introduce the concept of time-inversion referencing.
This is an extension of hypertext allowing authors to cite papers that where
not yet published (or even not yet written) when they publish a manuscript.
We are convinced that this concept will prove very useful especially when
adjustments to the so-called intellectual property rights have to be made.
We apply the concept to our paper on rotational properties of trapped bosons.
\end{abstract}

If the paper \cite{TIR} would have been published prior to Refs. \cite
{BLD9611090,BLDPRA97}, we would have added the reference \cite{TIR} in our
papers. Indeed, both \cite{TIR} and \cite{BLD9611090,BLDPRA97} deal with the
same problem (i.e. the moment of inertia of a confined Bose gas), the
calculations are done for the same model (bosons in a harmonic trap), with
the same method (projection on the symmetric representation of the
permutation group in combination with a Feynman-Kac functional), and lead to
the same final result (the expression for the moment of inertia in terms of
the system parameters). More methodological details can be found in Ref. 
\cite{BLDbasis}. Fortunately, we \cite{BLD9611090,BLDPRA97} also treated a
generalization to the case of {\em interacting} bosons using the
Jensen-Feynman theorem, which is not mentioned in \cite{TIR}. Therefore we
would not have had to abandon publication if we had foreseen the paper \cite
{TIR}.

We invite the casual reader to verify that a time-inversion reference is
appropriate here. It provides at least a warning against a substantially
incomplete bibliography. In extreme cases it might warrant intellectual
property rights.


\begin{references}
\bibitem{TIR}  J. Schneider and H. Wallis, {\sl Permutation cycles and the
moment of inertia of a trapped ideal Bose gas,} cond-mat-0003471 (2000).

\bibitem{BLD9611090}  F. Brosens, J. T. Devreese, and L. F. Lemmens, {\sl %
Rotational properties of trapped bosons,} cond-mat/9611090 (1996).

\bibitem{BLDPRA97}  F. Brosens, J. T. Devreese, and L. F. Lemmens, Phys.
Rev. A{\bf \ 55}, 2453 (1997).

\bibitem{BLDbasis}  F. Brosens, J. T. Devreese, and L. F. Lemmens, Phys.
Rev. E{\bf \ 55}, 227 (1997).
\end{references}
\end{document}